\documentclass[twocolumn,showpacs,preprintnumbers,amsmath,amssymb,superscriptaddress]{revtex4-1}

\usepackage{graphicx}
\usepackage{dcolumn}
\usepackage{bm}
\usepackage{graphics}
\usepackage{amsmath}
\usepackage{amssymb}
\usepackage{amscd}
\usepackage{bbm}
\usepackage{afterpage}
\usepackage{rotating}
\usepackage{fancyhdr}
\usepackage{epsfig}
\usepackage{theorem}
\usepackage{euscript}
\usepackage{hyperref}
\usepackage{subfig}
\begin{document}

\newcommand{\veps}{\varepsilon}
\newcommand{\lrarrow}{\leftrightarrow}
\newcommand{\beq}{\begin{equation}}
\newcommand{\eeq}{\end{equation}}
\newcommand{\bea}{\begin{eqnarray}}
\newcommand{\eea}{\end{eqnarray}}
\def\slash{\not\!}
\newcommand{\ud}{\textrm{d}}
\title{Exact constraints on D$\leq 10$ Myers-Perry black holes and the Wald Problem}
\author{Jason~Doukas}
\email[Email: ]{jasonad@yukawa.kyoto-u.ac.jp}
\affiliation{Yukawa Institute for Theoretical Physics, Kyoto University, Kyoto, 606-8502, Japan}

\begin{abstract}
  Exact relations on the existence of event horizons of Myers-Perry black holes are obtained in $D\leq 10$ dimensions. It is further shown that naked singularities can not be produced by ``spinning-up'' these black holes by shooting particles into their $\lfloor\frac{D-1}{2}\rfloor$ equatorial planes.  
\end{abstract}
\date{\today}
\preprint{YITP-10-86}
\maketitle
\section{Introduction}\label{sec:intro}
Singularities can be found in many physical theories. Generally speaking, rather than accepting the singularities as legitimate physical quantities we usually interpret them as non-physical solutions or a breakdown of the theory itself. In black hole spacetimes various types of singularities can exist; there are point-like structures found inside Schwarzschild solutions, ring singularities found inside Kerr black holes, and even more unusual topologies can arise in higher dimensions \cite{Myers1986304,PhysRevLett.88.101101}.

When these singularities are concealed by event horizons they can not causally interact with outside observers and their existence is of little consequence. 
However, a surprising feature of the Kerr solution \cite{PhysRevLett.11.237} is that for certain values of mass and angular momentum it describes a naked singularity, that is a singularity unshielded by any horizon.
Specifically, the uncharged solutions possess a horizon when
\begin{equation}
\Delta=r^2+a^2-2M r=0,
\end{equation}
where $r$ is the radial coordinate, $a$ is the angular momentum per unit mass, and $M$ is the mass of the black hole. Solving this quadratic equation leads to the conclusion that naked singularities are present when $M< a$. Such solutions are usually dismissed by appealing to the cosmic censorship hypothesis \cite{Penrose2002}, which posits that no singularity can be visible from future null infinity.
Perhaps the best evidence that Kerr black holes can not be turned into naked singularities can be found in the intriguing results of Wald \cite{Wald1974548}. He considers an extremal black hole and tries to create a naked singularity by injecting particles with enough angular momentum to achieve the above relation. However, it is found that either the particle misses the black hole completely or is repelled by gravitational dipole-dipole type forces and so a naked singularity is never formed. 

In recent times the exploration of extra spatial dimensions has been a dominant theme in high energy physics. One reason for this is that extra dimensions are necessary in quantum theories of gravity like string theory, secondly they have been shown in various guises to provide a novel solution to the hierarchy problem \cite{ArkaniHamed1998263, *PhysRevLett.83.3370}. Since these theories typically operate when gravitational effects are strong, higher dimensional black holes are expected to play a key role in the experimental interrogation of these theories; in some cases observational signatures are even predicted at the LHC \cite{PhysRevD.65.056010,*PhysRevLett.87.161602}. In order to test these theories and aid experimental searches it is useful to have exact constraints on the black hole parameters.

In higher dimensions uncharged black hole solutions with angular momentum have been found by Myers and Perry \cite{Myers1986304}.  To date a systematic study of the horizon constraints on these black holes has not been performed (although the qualitative features have been inferred in \cite{lrr-2008-6} and some results in the special case of $n$ equal non-zero spins can be found in the recent work \cite{PhysRevD.73.024017}). This is possibly due to the difficulty in finding algebraically closed expressions to the equation $\Delta=0$. We report in this letter that such constraints can in fact be determined exactly in all dimensions less than or equal to ten without having to solve this equation. 

Furthermore, we will show in section \ref{sec:nospinup} that these constraints allow us to analyze whether or not rotating black holes can be spun up into naked singularities in higher dimensions. Recently an effort was made to repeat the Wald gedanken experiments for Myers-Perry black holes \cite{PhysRevD.81.084051}. While their conclusions were that black holes could not be spun up into naked singularities, such analysis could only be performed for singly rotating black holes or those with all angular momenta equal. Even then, for dimensions greater than five, numerical methods were required. In this letter we go one step further and show that none of the Myers-Perry black holes can be spun up into naked singularities. This is done by generalising the setup of Wald to $d$ particles with arbitrary angular momentum and energy where a single particle is taken to fall into the black hole along each of the equatorial planes. 

This paper is broken up into three sections; first we describe the MP metric then we find the black hole mass and angular momentum constraints in section \ref{sec:AMconstraints}. In section \ref{sec:nospinup} we repeat the Wald experiment in the higher dimensional space and show that naked singularities can not be created by classical processes once a MP black hole is formed. 
\section{The MP metric}
Spinning black holes in $D$ spacetime dimensions, $\eta_{\mu\nu}=\text{diag}(-1,+1,\cdots,+1)$, become somewhat more complicated compared with their 4 dimensional counterparts. Due to the symmetries of flat space an independent angular momentum variable $a_i$ exists for each of the $\lfloor \frac{D-1}{2} \rfloor$ Casimirs of the little group of $SO(D-1)$. Since the metrics have slightly different forms for odd and even dimensions it will be convenient to define the variables $d$ and $\sigma$:
 \begin{equation}
d=
\begin{cases}
  \frac{D-2}{2}, & \mbox{if }D\mbox{ is even}\\
  \frac{D-1}{2}, & \mbox{if }D\mbox{ is odd}.\\
\end{cases}
\quad
\sigma=
\begin{cases}
  1, & \mbox{if }D\mbox{ is even}\\
  2 & \mbox{if }D\mbox{ is odd}.\\
\end{cases}
 \end{equation}
Note that $d$ is just the number of angular momentum parameters. Then the MP metric can be written:
\begin{eqnarray}\label{eqn:metric}
  ds^2&=&-dt^2 +(2-\sigma) r^2 d\alpha^2+\sum_{i=1}^d(r^2+a_i^2)(d\mu_i^2+\mu_i^2 d\phi_i^2)\nonumber\\
  &+&\frac{(\Pi-\Delta)}{\Pi F} \left(dt - \sum_{i=1}^da_i\mu_i^2 d\phi_i\right)^2+\frac{\Pi F}{\Delta} dr^2,
\end{eqnarray}
where\footnote{Note we have changed $a_i\leftrightarrow-a_i$ so that in the 4D limit the angular momentum in the z-direction corresponds to positive $a$. Furthermore, we use the mass parameter convention $\mu=2M$ so that in 4D $M$ agrees (with $G=1$) with the usual Kerr mass parameter.},
\begin{eqnarray}
1&=&  \sum_{i=1}^d\mu_i^2+(2-\sigma)\alpha^2,\quad F\equiv1-\sum_{i=1}^d\frac{a_i^2\mu_i^2}{r^2+a_i^2},\nonumber\\
  \Delta&\equiv&\Pi-2 M r^{\sigma}, \quad \Pi\equiv\prod_{i=1}^d(r^2+a_i^2).
\end{eqnarray}
From these metrics several other important properties can be established \cite{ Myers1986304} which we mention here for later use. Firstly the mass and angular momenta of the black hole are given by:
\begin{eqnarray}\label{eqn:aJrelation}
  \mathcal{M}& =&\frac{(D-2) A_{D-2} }{8 \pi G}M,\\
  J^{x^i y^i}& =&\frac{2}{D-2} \mathcal{M} a_i,
\end{eqnarray}
where $A_{D-2}$ is the area of the $(D-2)$-sphere, $A_{D-2}=2\pi^{(D-1)/2}\Gamma\left((D-1)/2\right)^{-1}$, and $G$ is Newton's constant. Since the mass parameter $M$ is proportional to the mass, $\mathcal{M}$, we will often refer to $M$ simply as the mass.

The angular velocity of the horizon in the plane defined by $\mu_i=1$ is given by
\begin{equation}
  \omega_i=\frac{a_i}{r_h^2+a_i^2},
\end{equation}
where $r_h$ is the radius of the outer horizon. The area of the horizon as originally presented \cite{Myers1986304} is:
\begin{equation}\label{eqn:area}
  \mathcal{A}=\frac{A_{D-2} M}{\kappa}\left(D-3-2\sum_{i=1}^d \frac{a_i^2}{r_h^2+a_i^2}\right),
\end{equation}
where $\kappa$ is the surface gravity:
\begin{equation}
  \kappa=\frac{\partial_r \Delta(r_h)}{4 M r_h^{\sigma}}.
\end{equation}
However, we note that $\Pi$ satisfies the differential equation:
\begin{equation}\label{eqn:pidiff}
 r\partial_r \Pi+ \sum_{i=1}^d a_i \partial_{a_i} \Pi=2d \Pi,
\end{equation}
which can be used to write the area in the more convenient form:
\begin{equation}\label{eqn:areasimp}
  \mathcal{A}=2M A_{D-2} r_h.
\end{equation}
Since the  location of the horizons are slightly different for odd and even dimensions it will be convenient in the next section to analyze these cases separately. 
\section{Mass and angular momentum relations}\label{sec:AMconstraints}
\subsection{$D$-even}
 In $D$-even dimensions the condition for the location of the horizon is:\\
\begin{equation}\label{eqn:devenhor}
  \Delta=\Pi-2 M r=0, \quad  \Pi=\prod_{i=1}^{d}(r^2+a_i^2).
\end{equation}
In their original work Myers and Perry pointed out that equation (\ref{eqn:devenhor}) is a polynomial in $r$ of order $(D-2)$, and therefore algebraically closed solutions could only be guaranteed for $D=4$ and $D=6$. They further noted that the polynomial is not completely generic so solutions for larger $D$ might be possible, but to the present authors knowledge in the general case no such results have been reported.

\begin{figure}[h]
  \centering
  \includegraphics[scale=.8]{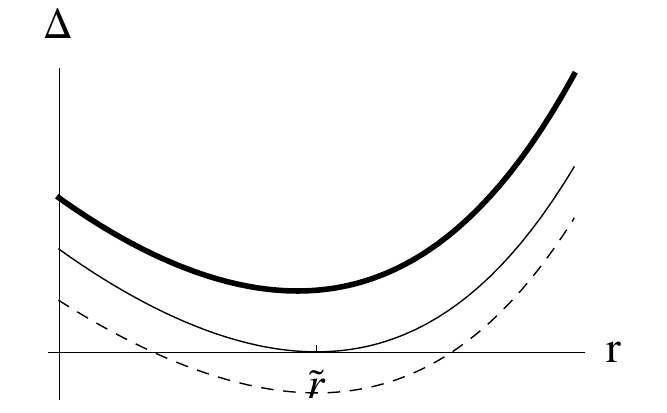}
  \caption{Thick line: $\Delta (\tilde{r})>0$ there is a naked singularity. Thin line: the minimum occurs at the $r$-intercept there is a degenerate horizon. Dashed line: $\Delta (\tilde{r})<0$ horizons occurs at $r$-intercepts.}
  \label{fig:scenarios}
\end{figure}

The method we take, while different to the above, makes use of many of the same arguments presented in \cite{Myers1986304}. The necessary observations are that for all $r>0$  $\partial_r \Pi$ is a strictly increasing function that is zero at $r=0$. Since $M>0$ there is exactly one real and positive minima, $\tilde{r}$, of $\Delta$: $\partial_r\Delta(\tilde{r})=0$. One can further show that $\Delta(r=0)\geq0$ (where equality occurs iff one or more of the spin parameters is zero) and $\partial_r\Delta(r=0)<0$. Therefore
there are only three possibilities for the number of horizons, depicted in figure \ref{fig:scenarios}; either $\Delta$ never crosses the axis leaving no solutions, just touches the axis in a degenerate solution or completely crosses the axis giving two distinct horizons.

Our strategy to find the allowed black hole parameter space proceeds as follows.
Let $\tilde{r}$ be the unique minima of $\Delta$ for positive $r$. Then the black hole will have a horizon if:
\begin{equation}\label{eqn:boundhorizon}
  \Delta (\tilde{r})\leq0,\quad \tilde{r}>0.
\end{equation}
The reason for restricting, $\tilde{r}$, to positive values is that when $\Delta(\tilde{r}=0)=0$, then the horizon at $\tilde{r}$ will have zero area (see equation (\ref{eqn:areasimp})) and so in this case a naked singularity would be exposed. Therefore strict inequality must be assumed in this case (i.e., when one or more of the spin parameters is zero). From here on this will be implied in our definition of the inequality symbol. 

Since the inequality is saturated when $\Delta(\tilde{r})=0$, equality occurs iff $\tilde{r}$ is also an $r$-intercept. Therefore,
\begin{equation}\label{eqn:evenconstraint}
  M\geq\frac{\Pi(\tilde{r}_h)}{2\tilde{r_h}},
\end{equation}
where $\tilde{r}_h$ is defined as a solution to the simultaneous set of equations
\begin{eqnarray} \label{eqn:constr1}
  \Delta(\tilde{r}_h)=0,\quad \partial_r \Delta (\tilde{r}_h)= 0.
\end{eqnarray}
The solution to these equations defines a surface in the parameter space for which the horizons are degenerate. 
Since we have two constraint equations we can find $\tilde{M}(\{a_i\})$ and $\tilde{r}_h(\{a_i\})$ solely in terms of the angular momentum parameters, where $\tilde{M}$ and $\tilde{r}_h$ are the extremal mass and radius for a given set of $\{a_i\}$.

It is important to emphasize that finding the degenerate subspace turns out to be significantly easier than solving equation (\ref{eqn:devenhor}). Furthermore,  once $\tilde{r}_h$ is found it is a relatively straightforward task to input it into equation (\ref{eqn:evenconstraint}), and thereby obtain the constraint. Eliminating $M$ from equations (\ref{eqn:constr1}) results in the single equation:
\begin{equation}\label{eqn:evenequation}
  0=\tilde{r}_h\partial_r \Pi (\tilde{r}_h)-\Pi(\tilde{r}_h).
\end{equation}
Using the product expansion:
\begin{equation}
  \prod_{i=1}^{j}(r^2+a_i^2)=\sum_{i=0}^{j}r^{2 i} A_{j}^{j-i},
\end{equation}
with the coefficients conveniently defined as
\begin{equation}
  A^k_n=\sum_{\nu_1<\nu_2<\dots\nu_k}a_{\nu_1}^2a_{\nu_2}^2\dots a_{\nu_k}^2,
\end{equation}
where $v_k$ are summed over the range $[1,n]$, $0\leq k\leq n$ and $A^0_k\equiv 1$, equation (\ref{eqn:evenequation}) can be written as
\begin{equation}
  P_2(\tilde{r}_h)=\sum_{i=0}^{d}(2 i-1)\tilde{r}_h^{2 i} A_{d}^{d-i}=0.
\end{equation}
We observe that upon redefining $\tilde{r}_h$ in terms of a squared variable, the equation has algebraically closed solutions for all even $D\leq10$.

This polynomial can also be used to find an upper bound on the extremal radius, which may be useful for example in larger than ten dimensions. Since $P_2$ is less than zero at the origin and equal to zero at the extremal horizon an upper bound on $\tilde{r}_h$ can be determined by finding a value of $r$ for which the polynomial is positive. Since it is only the constant term which is negative it is easy to see, 
\begin{equation}
  \tilde{r}_h\leq \min_i{|a_i|}.
\end{equation}

As an example of how the polynomial can be used to find an exact relation, we calculate the constraint in $D=6$. In this case the polynomial reads:
\begin{equation}
  0=3 \tilde{r}^4_h+(a_1^2+a_2^2)\tilde{r}^2_h-a_1^2a_2^2,
\end{equation}
with solution given by
\begin{equation}
  \tilde{r}^2_h=\frac{-(a_1^2+a_2^2)+\sqrt{(a_1^2+a_2^2)^2+12 a_1^2 a2^2}}{6},
\end{equation}
so the $D=6$ Myers-Perry black hole has a horizon iff
\begin{equation}
  M\geq \frac{(\tilde{r}^2_h+a_1^2)(\tilde{r}^2_h+a_2^2)}{2\tilde{r}_h}.
\end{equation}
It is satisfying to note that if we put $a_2=0$ the constraint becomes $M>0$ reproducing the well known result that in even dimensions there is always a horizon if one or more of the spin parameters is zero \cite{Myers1986304}.

In figures \ref{fig:D6space} and \ref{fig:solidD8} we plot the allowed parameter space for black holes normalised with $M=1$ for $D=6$ and $ D=8$ respectively. These figures nicely agree with the qualitative features predicted by the technique described in \cite{lrr-2008-6}.

\subsection{$D$-odd}
Next we consider the $D$-odd case. In this case, it is more convenient to use $l=r^2$. The location of the horizon is found by the equation:
\begin{equation}\label{eqn:Deltal}
\Delta=\Pi(l)-2M l=0, \quad   \Pi=\prod_{i=1}^{d}(l+a_i^2).
\end{equation}
As pointed out by Myers and Perry, by Galois' theorem this equation has closed solutions for $D=5,7,9$ (and even a closed solution in terms of elliptic functions for $D=11$). Our method unfortunately does not get any further in the number of dimensions. However, it does allow the constraint to be easily determined.

Starting from equation (\ref{eqn:Deltal}), as in the even case $\partial_l\Pi$ is strictly increasing for $l>0$, however now $\partial_l\Pi(l=0)$ can be greater than zero. This implies $\Delta$ will have a unique minimum for $l>0$ iff $2 M > \partial_l \Pi(l=0)$. First consider the solutions for which $2 M \leq \partial_l \Pi(l=0)$. Because $\Delta(l=0)\geq0$ and $\Delta(l\rightarrow\infty)=+\infty$ and because there are no turning points for $l>0$ the only position a horizon could occur at is $l=0$, but this would be a zero area horizon and would therefore expose a singularity. Therefore, we must have  $2M>\partial_l\Pi(l=0)$. Since $\Delta(l=0)\geq0$ we find that we have the same possibilities for the number of horizons as in the even case. Performing the same analysis again we find that:
 \begin{equation}\label{eqn:leqn}
  \tilde{l}_h\partial_l\Pi(\tilde{l}_h)-\Pi(\tilde{l}_h)=0, \quad M\geq\frac{\Pi(\tilde{l}_h)}{2\tilde{l}_h}.
\end{equation}
Where again the inequality in equation (\ref{eqn:leqn}) must be replaced by a strict inequality when $\tilde{l}_h=0$. The polynomial for the odd case is:
\begin{equation}
  P_1(\tilde{r}_h)= \sum_{i=0}^{d}(i-1) \tilde{l}_h^{i} A_{d}^{d-i}=0.
\end{equation}

Using a similar argument to the one used in the even case we can use this polynomial to find an upper bound on the extremal radius:
\begin{equation}
  \tilde{r}_h\leq \min_{ i\neq j } \{\sqrt{a_i a_j}\}. 
\end{equation}

It is easy to verify that for $D=5$ the polynomial becomes $\tilde{l}_h^2=a_1^2a_2^2$ and we recover the known relation:
\begin{equation}
2 M\geq 2 |a_1 a_2|+a_1^2+a_2^2.
\end{equation}
When $D=7$ we have:
\begin{eqnarray}
\label{eqn:p}
  \tilde{l}_h&=&\frac{1}{6}\left(-(a_1^2+a_2^2+a_3^2)+\frac{1}{p}(a_1^2+a_2^2+a_3^2)^2+p\right), \nonumber\\
  p^3&=&-(a_1^2+a_2^2+a_3^2)^3+54 a_1^2 a_2^2 a_3^2+ 6\sqrt{3}\\
  &\times& \sqrt{-(a_1^2+a_2^2+a_3^2)^3 a_1^2a_2^2a_3^2+27 (a_1^4a_2^4 a_3^4)},\nonumber
\end{eqnarray}
and
\begin{equation}
  M\geq\frac{(\tilde{l}_h+a_1^2)(\tilde{l}_h+a_2^2)(\tilde{l}_h+a_3^2)}{2\tilde{l}_h}.
\end{equation}
Equation (\ref{eqn:p}) actually defines three solutions; the cube root must be taken so that $\tilde{l}_h$ is non-negative. 
When we set one spin to zero, we get the expected exceptional behaviour $\tilde{l}_h=0$ and the bound is replaced by a strict inequality. In fact, in any odd dimension for $a_1=0$:
\begin{equation}\label{eqn:onezerobound}
  M> \frac{1}{2}\prod_{i=2}^{d} a_i^2. 
\end{equation}
Furthermore, setting one more spin to zero gives the usual unbounded behaviour \cite{Myers1986304}.

\begin{figure}[t]
\centering
\subfloat[$D=7$]{\label{fig:solidD7} \includegraphics[scale=.8]{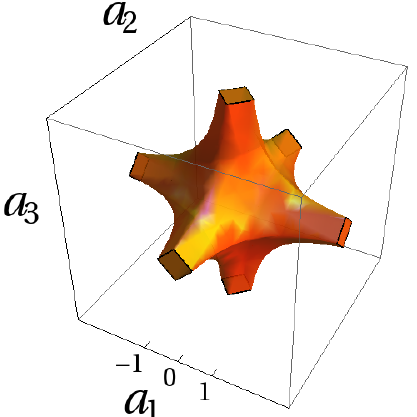} }
\subfloat[$D=8$]{\label{fig:solidD8} \includegraphics[scale=.8]{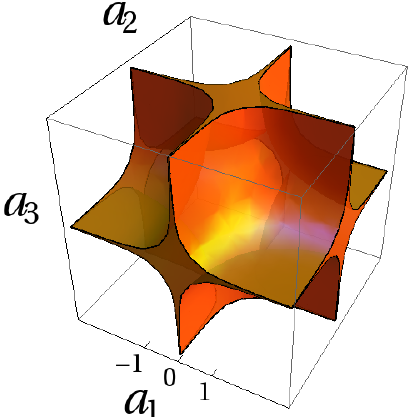} }
\caption{The solid region indicates parameters for which black holes have a horizon. $M=1$. (left) D=7, the prongs extended out towards infinity, in the $a_i=0$ plane the remaining two spins are bounded by hyperbolas (\ref{eqn:onezerobound}). (right) D=8, in this case the $a_i=0$ planes are always part of the allowed solutions. }  \label{fig:solids}
\end{figure}

In figures \ref{fig:D5space} and \ref{fig:solidD7} we plot the black hole parameter space ($M=1$) for $D=5$ and $ D=7$ dimensions respectively. This again validates the qualitative features predicted by the technique described in \cite{lrr-2008-6}.

\subsection{Other MP varieties} 
Since solutions with the full set of non-zero angular momentum parameters can be difficult to examine, several varieties of these black holes are commonly explored. The most important of these being those with all spins equal, these are also known as cohomogeneity-1 black holes. We find for these that:
\begin{equation}
  2M\geq 
\begin{cases}
  (2d)^d|a|^{(2d-1)}(2d-1)^{1/2 -d}, & \mbox{if }D\mbox{ is even}\\
  a^{2d-2}d^d(d-1)^{1-d}, & \mbox{if }D\mbox{ is odd},\\
\end{cases}
\end{equation}
in agreement with what has been presented previously in the literature \cite{Nozawa2005}.

It is interesting to consider the large dimension limit of these solutions. When the number of angular momentum parameters, $d\rightarrow\infty$, the solutions will have a horizon for $a<1$, regardless of the value of $M>0$. In the same limit, for $a\geq1$ the mass would need to be infinitely large to avert naked singularities.

\begin{figure}
  \centering
  \subfloat[$D=5$; $a=a_1$, $b=a_2$]
  {\label{fig:D5space}\includegraphics{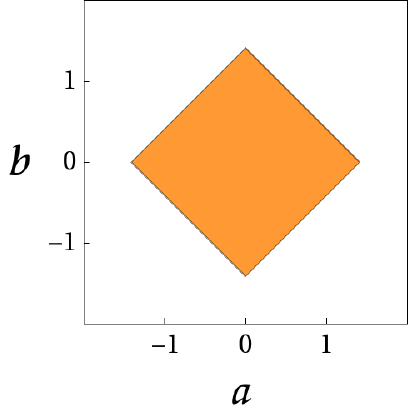}}
  \subfloat[$D=6$; $a=a_1$, $b=a_2$]
  {\label{fig:D6space}\includegraphics{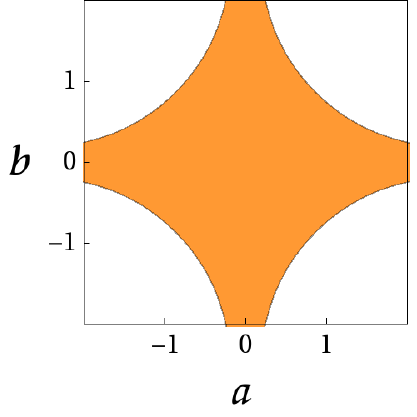}}\\
  \subfloat[$D=7$; $a=a_1=a_3$, $b=a_2$]
  {\label{fig:D7space}\includegraphics{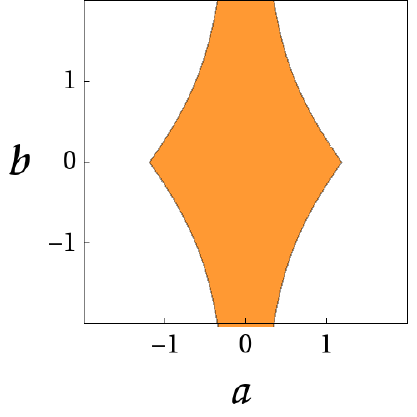}}
  \subfloat[$D=8$; $a=a_1=a_3$, $b=a_2$]
  {\label{fig:D8space}\includegraphics{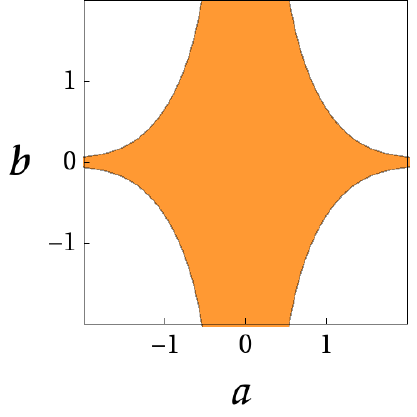}}\\
   \subfloat[$D=9$; $a=a_1=a_3$, $b=a_2=a_4$]
  {\label{fig:D9space}\includegraphics{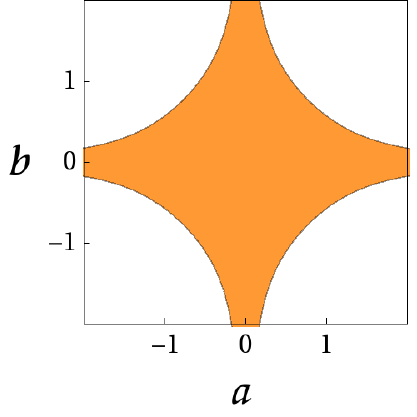}}
  \subfloat[$D=10$; $a=a_1=a_3$, $b=a_2=a_4$]
  {\label{fig:D10space}\includegraphics{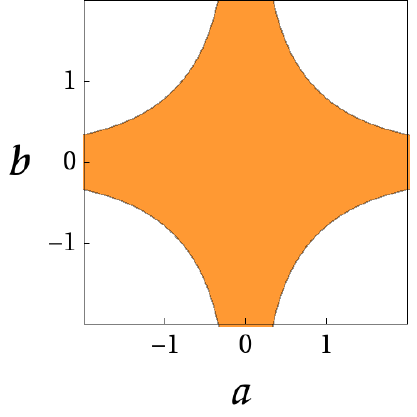}}
  \caption{Cohomogeneity-2 black hole regions $M=1$. Plots (a) and (b) are the usual $D=5$ and $D=6$ dimensional black hole regions with a unique parameter on each axis. In the other cases some of the parameters have been equated as shown in the subtext. }
  \label{fig:com2}
\end{figure}

\begin{figure}
  \centering
  \subfloat[$D=9$; $a=a_1=a_4$, $b=a_2$, $c=a_3$]
  {\label{fig:D9C3space}\includegraphics{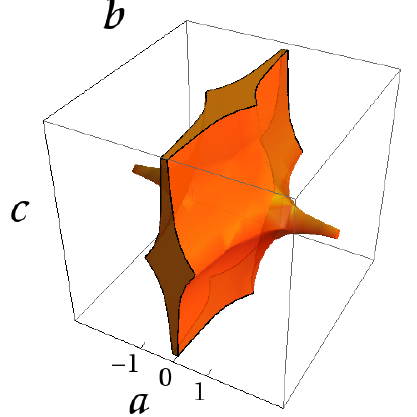}}
  \subfloat[$D=10$; $a=a_1=a_4$, $b=a_2$, $c=a_3$]
  {\label{fig:D10C3space}\includegraphics{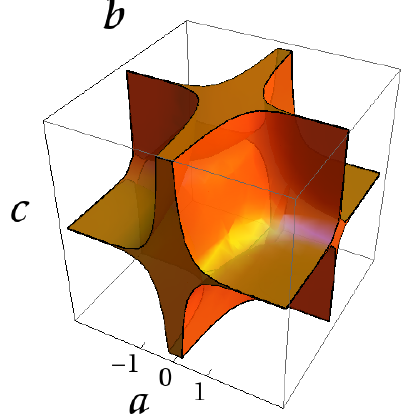}}
  \caption{Cohomogeneity-3 black hole solids $M=1$.}
  \label{fig:com3}
\end{figure}

While equating all the spin parameters greatly simplifies the solutions much of the interesting behaviour is lost in this process. However, one can generalise this idea to higher cohomogeneity so that some of the features of multiple parameters can be probed while simultaneously keeping a lot of symmetry. Cohomogeneity-2 black holes are constructed by taking the whole set of non-zero spin parameters and setting each one of them to either the value $a$ or the value $b$. Plots of the angular momenta constraints for cohomogeneity-2 black holes can be seen in figure \ref{fig:com2}. Likewise the process can be generalised to cohomogeneity-3 black holes where three sets of parameters $a$,$b$ and $c$ are used, we show the cohomogeneity-3 black holes in dimensions $D=9$ and $D=10$ in figure \ref{fig:com3}.

Next we will use the constraints obtained to show, using arguments similar to that of Wald, that in higher dimensions MP black holes can not be spun-up past the extremal solution. 
\section{Classically spinning up a Myers-Perry black hole}\label{sec:nospinup}
From our analysis of the horizon structure we were able to determine the condition on $M$ and $a_i$ under which the black hole was an extremal solution. Having these relations are equivalent to having in the 4D case the well known equation $M^2=a^2+e^2$. It is then natural to see whether the Wald analysis, for which the extremal constraint equation was critical, can be carried out in higher dimensions. In effect, starting with an extremal solution, we want to see whether we can destroy the horizon by throwing in particles with orbital angular momentum.

We consider $d$ particles falling along each of the $d$-equatorial planes \footnote{We only need to consider motion along the equatorial place since the transfer of orbital angular momentum from the particle to the black hole will be greatest along this direction.}. Each particle is given an independent energy $E_i$ and angular momentum $L_i$.

From our previous results the extremal solution was given by:
\begin{equation}
  \tilde{M}=\frac{\Pi(\tilde{r}_h,a_i)}{2 \tilde{r}_h^{\sigma}}.
\end{equation}
Using equations (\ref{eqn:evenequation}) or (\ref{eqn:leqn}) it can be shown that $\partial_{\tilde{r}_h}\tilde{M}=0$ therefore, as we expect, the extremal mass is only a function of the spin parameters and we can write:
\begin{eqnarray}
  d\tilde{M}= \sum_{i=1}^d\frac{\partial \tilde{M}}{\partial a_i} da_i=\frac{1}{2 \tilde{r}_h^{\sigma}} \sum_{i=1}^d a_i \frac{\partial \Pi}{\partial a_i} \frac{d a_i}{a_i}, 
\end{eqnarray}
The spin parameter is a function of both the angular momentum in the $i$-th direction and the mass (\ref{eqn:aJrelation}), therefore, the spin parameter will change when there is an increase in the mass of the black hole even if there is no angular momentum absorbed in that direction. Since $d \tilde{M}=E_T$ where $E_T$ is defined as the sum of all the energies $E_i$ we obtain:
\begin{equation}\label{eqn:deltaa}
  \frac{da_i}{a_i}=\frac{1}{\tilde{\mathcal{M}}}\left(\frac{(D-2)}{2}\frac{L_i}{a_i}-E_T\right).
\end{equation}
First we assume that none of the spins are zero. At the extremal radius equation (\ref{eqn:pidiff}) reduces to the relationship:
\begin{equation}
  \sum_{i=1}^d a_i \frac{\partial \Pi(\tilde{r}_h)}{\partial a_i}=(D-3)\Pi(\tilde{r}_h),
\end{equation}
Using this we find that for the particles to create a naked singularity the total energy absorbed by the black hole must be less than:
\begin{equation}
  E_T<\sum_{i=1}^d\frac{a_i L_i}{\tilde{r}_h^2+a_i^2}=\sum_{i=1}^d\omega_i L_i.
\end{equation}
If one or more of the spin parameters are zero then $\tilde{r}_h=0$. In this case there is no extremal solution since the horizon area vanishes. Nevertheless, our approach would still be valid for arbitrarily small but non-zero values of one or more spin parameters. 

We now investigate how much energy the infalling particles can have at the horizon. Since all $d$ equatorial planes are symmetrical, we only need to consider the geodesic of a single particle in the metric defined along the $\mu_1=1$ equatorial plane \cite{Nozawa2005}; the following analysis will be identical for each of the other infalling particles. In this plane the metric (\ref{eqn:metric}) becomes:
\begin{eqnarray}
  ds^2_1&=&-\left(1-\frac{(\Pi-\Delta)(r^2+a_1^2)}{\Pi r^2}\right)dt^2+\frac{\Pi r^2}{\Delta(r^2+a_1^2)}dr^2\nonumber\\&+&\left(r^2+a_1^2+ \frac{a_1^2(\Pi-\Delta)(r^2+a_1^2)}{\Pi r^2}\right) d\phi_1^2\nonumber\\
  &-&\frac{2a_1(\Pi-\Delta)(r^2+a_1^2)}{\Pi r^2}dt d\phi_1.
\end{eqnarray}
The geodesic equation \cite{Wald1974548,Cardoso2009} is given by
\begin{equation}
  \frac{D^2x^{\mu}}{Ds^2}=\frac{d^2x^{\mu}}{d s^2} +\Gamma^{\mu}_{\rho\sigma}\frac{dx^\rho}{ds}\frac{dx^\sigma}{ds}=0,
\end{equation}
which can be derived from the Lagrangian
\begin{eqnarray}
  \mathcal{L}&=&\frac{1}{2}g_{\mu\nu}(x)\frac{dx^\mu}{ds}\frac{dx^{\nu}}{ds},\nonumber\\
  &=&\frac{1}{2}(g_{tt} \dot{t}^2+g_{\phi\phi} \dot{\phi}^2+2 g_{t\phi} \dot{t}\dot{\phi}+g_{rr} \dot{r}^2).
\end{eqnarray}
Since the Lagrangian is independent of $t$ and $\phi$ there are two conserved quantities along the particle worldline which are associated with the energy and angular momentum at infinity respectively:
\begin{eqnarray}
-E&=&p_t=\frac{\partial \mathcal{L}}{\partial\dot{t}}=g_{tt} \dot{t}+g_{t\phi}\dot{\phi},\\
L&=&p_{\phi}=\frac{\partial\mathcal{L}}{\partial{\dot{\phi}}}=g_{\phi\phi}\dot{\phi}+g_{t\phi}\dot{t}.
\end{eqnarray}
We consider both time-like ($\delta=-1$) and null ($\delta=0$ )  trajectories. Then from the particle 4-velocity we have:
\begin{eqnarray}
  \delta = g_{\mu\nu} \frac{dx^{\mu}}{ds} \frac{dx^{\nu}}{ds}=2\mathcal{L}.
\end{eqnarray}
Given that:
\begin{equation}
  g_{t\phi}^2-g_{\phi\phi}g_{tt}=\frac{\Delta r^2}{\Pi F}\equiv \mathbb{g}
\end{equation}
it is relatively straight-forward to invert the metric and find that along the trajectory $E$ satisfies the equation:
\begin{equation}
  0=g_{\phi\phi} E^2+2 LE g_{t\phi}+g_{tt}L^2+(\delta-g_{rr}\dot{r}^2)\mathbb{g} 
\end{equation}
Solving this quadratic equation, keeping the solution for which the energy at infinity is positive, we obtain:
\begin{equation}
  E_i\geq - \frac{L_i g_{t\phi}}{g_{\phi \phi}}.
\end{equation}
The relevant metric components at the horizon are: 
\begin{eqnarray}
  g_{t\phi}=-\frac{a_1 (\tilde{r}^2_h+a_1^2)}{\tilde{r}^2_h},\quad  g_{\phi\phi}=\frac{(\tilde{r}^2_h+a_1^2)^2}{\tilde{r}_h^2}.
\end{eqnarray}
We are then able to deduce that the sum of the energies of all the $d$ particles at the horizon indeed satisfies:
\begin{equation}
  E_T\geq \sum_{i=1}^dL_i \omega_i,
\end{equation}
which is precisely the inequality necessary to prevent these processes from destroying the horizon. Therefore naked singularities can not be formed by spinning up MP black holes.

It is worth mentioning that even if a black hole satisfies the constraints presented it may not be stable; classical instabilities \cite{PhysRevLett.70.2837,*EmparanMyers,*Dias2009,*Dias2010} are known to arise in the ultra spinning regimes. In arriving at our main result in section \ref{sec:nospinup} classical particles obeying well-defined geodesic trajectories were used, it would be interesting to investigate whether the quantum nature of these particles alters our conclusions \cite{Matsas2007,*Hod2008}.

The author thanks W. Naylor for encouraging him to look for these constraints and H. T. Cho \& A. S. Cornell for discussions. The author also thanks D.Page who at the APCTP focus program on the frontiers of black hole physics, Pohang, noticed an inconsistency associated with an error in an earlier version of equation (\ref{eqn:deltaa}).  This work was supported by the Japan Society for the Promotion of Science (JSPS), under fellowship no. P09749.
\bibliography{MP_horizons}

\end{document}